\documentclass[a4paper,11pt]{article}
\pdfoutput=1 

\usepackage{jcappub} 
\usepackage{mypkgs}
\usepackage{xcolor}

\title{\boldmath Novel structures and collapse of solitons in nonminimally gravitating dark matter halos}

\author[a]{Jiajun Chen}
\author[b,*]{and Hong-Yi Zhang\note[*]{Corresponding author.}}

\affiliation[a]{School of Physical Science and Technology, Southwest University, Chongqing 400715, China}
\affiliation[b]{Tsung-Dao Lee Institute \& School of Physics and Astronomy, Shanghai Jiao Tong University, Shanghai 201210, China}

\emailAdd{chenjiajun@swu.edu.cn}
\emailAdd{hongyi18@sjtu.edu.cn}

\abstract{Ultralight dark matter simulations predict condensates with short-range correlation, known as solitons or boson stars, at the centers of dark matter halos. This paper investigates the formation and collapse of dark matter solitons influenced by nonminimal gravitational effects, characterized by gradient-dependent self-interactions of dark matter and an additional source in Poisson's equation for gravity. Our simulations suggest that the initial evolution of dark matter resembles that without nonminimal gravitational effects. However, regions with negative potential curvature may develop, and solitons will collapse when their densities reach certain critical values for both positive and negative coupling constants. With strong nonminimal gravitational effects, we verify that linear density perturbations could grow on both large and small scales, potentially enhancing structure formation.}

\begin{document}
\maketitle
\flushbottom

\section{Introduction}
A popular idea of dark matter involves a sub-eV scalar field known as ultralight dark matter \cite{Hui:2021tkt, Ferreira:2020fam, Matos:2023usa, OHare:2024nmr}. This class of models can produce cosmological structures similar to cold dark matter on large scales \cite{Hu:2000ke, Hui:2016ltb}. On small scales, the thermalization of scalar particles typically leads to the formation of condensations with short-range correlations called solitons or boson stars \cite{Guth:2014hsa, Zhang:2024bjo}. Focusing on the very light end of the mass spectrum $\cal O(10^{-22}$--$10^{-20}) \rm{eV}$, simulations in both expanding and nonexpanding backgrounds show that solitons emerge at the centers of dark matter halos \cite{Schive:2014dra, Levkov:2018kau, Veltmaat:2018dfz, Mocz:2019pyf, Eggemeier:2019jsu, May:2021wwp, Ellis:2022grh, Gorghetto:2022sue, PhysRevD.104.083022, Amin:2022pzv, PhysRevD.108.083021, Jain:2023ojg, Jain:2023tsr}. Solitons exhibit flat and smooth central density profiles, offering a natural solution to the discrepancy between the cuspy density profiles predicted by cold dark matter simulations and the flatter ones observed in galactic centers \cite{Ferreira:2020fam}. Ultralight dark matter is also proposed as a solution to the final parsec problem in simulations of supermassive black hole mergers \cite{Milosavljevic:2002ht}. In these simulations, inspirals stall at a few parsecs apart due to insufficient gravitational wave emission for orbits on those scales, which is problematic given the abundance of very massive black holes observed at galactic centers. However, orbiting near a solitonic core \cite{Koo:2023gfm} or interacting with ultralight dark matter granules \cite{Bromley:2023yfi} could provide mechanisms to facilitate orbital decay.\footnote{Other potential solutions to the final parsec problem involve, for example, self-interacting dark matter \cite{Alonso-Alvarez:2024gdz}, accretion disks \cite{Kocsis:2010xa, Goicovic:2016dul, Goicovic:2018xxi}, or axisymmetric galactic halo profile \cite{Khan:2013wbx} (which was debated in \cite{Vasiliev:2013nha}).}

Solitons give rise to important opportunities for discovering and constraining ultralight dark matter. For examples, solitons might dominate the bulge dynamics in massive galaxies and could potentially represent the majority of halo mass in smaller dwarf galaxies. Analysis against observed galactic kinematics has provided strong constraints on DM mass and self-interactions \cite{Schive:2014hza, Marsh:2015wka, Chen:2016unw, Gonzalez-Morales:2016yaf, Broadhurst:2019fsl, Bar:2018acw, DeMartino:2018zkx, Pozo:2023zmx, ParticleDataGroup:2022pth, Chakrabarti:2022owq, Dave:2023egr, Dave:2023wjq}. If dark matter is coupled to photons, solitons could generate observable electromagnetic signals \cite{Hertzberg:2018zte, Hertzberg:2020dbk, Amin:2020vja, Amin:2021tnq, Amin:2023imi}. 

Therefore, a detailed understanding about the formation and growth of solitons is crucial for inferring their cosmological abundance. In the simplest scenario, where dark matter interacts only through minimal couplings to gravity, solitons can condensate from a thermal bath of dark matter particles through kinetic relaxation \cite{Levkov:2017paj}. They continue to grow until relativistic effects become important. It is suggested that soliton growth slows once they become massive enough to heat the surrounding halo \cite{Eggemeier:2019jsu, PhysRevD.104.083022}, with the masses of the soliton $M_s$ and the halo $M_h$ approximately following the scaling $M_s \propto M_h^{1/3}$. However, there is ongoing debate about the quantitative characterization of the soliton growth rate \cite{Levkov:2018kau, Eggemeier:2019jsu, PhysRevD.104.083022, Chan:2022bkz, Dmitriev:2023ipv}. Baryonic matter could influence this process, typically leading to more compact cores \cite{Veltmaat:2019hou, Chan2018stars}.

As dense clumps of particles, dark matter densities within solitons could be many orders of magnitude higher than the local density $\sim 0.4\rm{GeV/cm^3}$ inferred from galactic rotation curves \cite{Bovy:2012tw, Pato:2015dua}. This implies that solitons are likely to be influenced by dark matter self-interactions and the foregoing simple picture needs to be improved. These self-interactions, to the leading order in the nonrelativistic limit, are quartic in fields (cubic terms are not allowed since a single nonrelativistic particle cannot create two nonrelativistic particles and vice versa). In the case of axions---hypothetical particles proposed to explain the absence of the neutron electric dipole moment \cite{Peccei:1977hh, Weinberg:1977ma, Wilczek:1977pj}---attractive self-interactions can destabilize solitons when their mass exceeds a threshold \cite{Chavanis:2011zi, Schiappacasse:2017ham, Visinelli:2017ooc, Zhang:2024bjo, Chavanis:2017loo, Chavanis:2022fvh, Zhang:2020bec, Zhang:2020ntm}, resulting in bursts of relativistic axions \cite{Levkov:2016rkk}. On the other hand, soliton condensation could be promoted by repulsive self-interactions \cite{PhysRevD.104.083022}, which can arise from spontaneous symmetry breaking and integrating out heavy degrees of freedom \cite{Fan:2016rda}. 

Dark matter self-interactions can also result from their nonminimal couplings to curvature fields such as $\xi\f^2 R/2$, where $\f$ is a scalar dark matter field, $R$ is the Ricci scalar, and $\xi$ characterizes the coupling strength \cite{Zhang:2023fhs, Zhang:2024bjo}.\footnote{Nonminimal couplings to gravity could arise from quantum corrections and are essential for the renormalization of field theories in curved spacetime \cite{Birrell:1982ix, Weinberg:1995mt, callan1970new, freedman1974energy, FREEDMAN1974354}. Their phenomenological implications have been studied in various contexts, such as dark matter \cite{Ivanov:2019iec, Ji:2021rrn, Sankharva:2021spi, Barman:2021qds, Zhang:2023fhs}, inflation \cite{Starobinsky:1979ty, Turner:1987bw, Ford:1989me, Faraoni:2000wk, Golovnev:2008cf, Golovnev:2008hv, Golovnev:2009ks, Golovnev:2009rm}, and modified gravity \cite{Moffat:2005si, Brownstein:2005zz, Tasinato:2014eka, Heisenberg:2014rta, DeFelice:2016yws, deFelice:2017paw}.} This leading-order nonminimal gravitational interaction (NGI) manifest as quartic terms in fields, including both gradient-independent and gradient-dependent parts. At first glance, the gradient-dependent part seems negligible due to the suppression by the small momenta of particles, reducing the effective self-interactions to the traditional parameterization used in previous references. However, as shown in \cite{Zhang:2024bjo}, these two components are interrelated because they arise from integrating out the nonlocal gravitational potential, described by a modified Poisson's equation. In this paper, we aim to investigate the impact of this new component of self-interactions on the formation and growth of solitons using 3+1-dimensional simulations. We will see that regions with negative mass density (defined as the source of the Poisson's equation) could develop. By tracking the evolution of maximum densities, we demonstrate that gradient-dependent self-interactions impose an upper bound on the mass of solitons for both positive and negative couplings, thereby affecting the mass distribution of solitons in the long-term evolution.

The rest of the paper is organized as follows. In section \ref{sec:model}, we introduce the key components of nonminimally gravitating dark matter and outline our numerical strategy. The results of our simulations are presented in section \ref{sec:formation}. First, we reproduce the numerical evolution of dark matter without NGIs in section \ref{sec:no_ngi}. Next, we discuss key properties of nonminimally gravitating solitons based on theoretical arguments in \ref{sec:soliton_ngi}. The evolution of nonminimally gravitating dark matter with positive and negative couplings is analyzed in sections \ref{sec:pos_ngi} and \ref{sec:neg_ngi}, respectively. Finally, we summarize our results in section \ref{sec:conclusion}.

\section{Field equations and numerical setup}
\label{sec:model}
The nonrelativistic dynamics of ultralight dark matter can be described by the Schroedinger equation coupled to gravity through Poisson's equation \cite{Hui:2021tkt, Ferreira:2020fam, Matos:2023usa, OHare:2024nmr, Salehian:2021khb}. When incorporating both minimal and nonminimal couplings to gravity, the Schroedinger-Poisson equations for scalar dark matter are \cite{Zhang:2024bjo}
\begin{align}
\label{eom1}
i\pd_t\psi &= -\frac{\nabla^2}{2m}\psi + m\Phi \psi + \frac{\xi}{2 m\MP^2} \rho_\xi \psi ~,\\
\label{eom2}
\nabla^2\Phi &= \frac{1}{2\MP^2} \rho_\xi ~,
\end{align}
where $\psi$ is a nonrelativistic classical field that varies slowly in time, $m$ is the field mass, $\Phi$ is the gravitational potential, $\xi$ characterizes NGIs, and $\rho_\xi$ is the mass density defined as $\rho_\xi \equiv \rho+ (\xi/m^2)\nabla^2\rho$ with $\rho\equiv m|\psi|^2$. The natural units $c=\hbar=1$ are adopted. In this work, we focus on time scales short compared with the Hubble time, thus neglecting the expansion of the universe.

Unlike the usual equations used in the literature \cite{Hui:2021tkt, Ferreira:2020fam, Matos:2023usa, OHare:2024nmr}, Poisson's equation here is sourced by $\rho_\xi$ rather than the rest mass density $\rho$ due to nonminimal gravitational effects. The NGI term in \eqref{eom1} has two components: one resembles the self-interaction from a quartic scalar dark matter coupling, and the other depends on the density gradient. As shown in \cite{Zhang:2024bjo}, the gradient-dependent part imposes an upper limit on the density hence mass of dark matter solitons, independent of the sign of $\xi$. We will discuss it in more details in section \ref{sec:soliton_ngi}.

The equations \eqref{eom1} and \eqref{eom2} can be simplified by introducing the dimensionless quantities (denoted with tilde):
\begin{align}
\label{nondimensionalize}
t \rightarrow \frac{\til t }{v^2 m}\sep
\b x \rightarrow \frac{\til{\b x}}{v m} \sep
\Phi \rightarrow v^2 \til\Phi \sep
\psi_i \rightarrow v^2 \sqrt{m} F \til \psi_i \sep
\xi \rightarrow \frac{\til\xi}{v^2} ~,
\end{align}
where $v$ is an arbitrary positive constant and $F$ is a chosen mass scale, which we set $F = \sqrt{2} \MP$. As a result, equations \eqref{eom1} and \eqref{eom2} transform to
\begin{align}
\label{eom3}
i \pd_{\til t} \til\psi &= - \frac{\til\nabla^2}{2} \til\psi + \til\Phi\til\psi + \til\xi \til\rho_\xi \til\psi ~,\\
\label{eom4}
\til\nabla^2\til\Phi &= \til\rho_\xi ~,
\end{align}
where the mass density $\til\rho_\xi = \til\rho + \til\xi \til\nabla^2\til\rho$ and the particle number density $\til\rho = |\til\psi|^2$. These equations are independent of $v$, implying a scaling symmetry that allows us to set $|\til\xi|=1$. However, to manage numerical challenges associated with small-scale structures and long-term evolution, we will retain flexibility in choosing different values for $\til\xi$.

To simulate the Schroedinger-Poisson equations \eqref{eom3} and \eqref{eom4}, we employ a fourth-order pseudospectral method as described in \cite{Du:2018qor,Chen:2024vgh,Chen:2021nnf}, treating the NGI term in the Schroedinger equation as an additional part of the potential. Initial conditions are set in momentum space with a delta-like distribution $|\til \psi_{\til{\b p}}|^2 \propto \delta (|\til{\b p}| - v_0/v)$ and a random phase. Here, $v_0\ll 1$ is a reference velocity for the nonrelativistic system, but its exact value is unimportant due to the freedom in choosing $v$. For simplicity, we set $v=v_0$. Alternative initial conditions, such as a Gaussian distribution $|\til\psi_{\til{\b p}}|^2 \propto e^{-\til{\b p}^2}$ or a step-like distribution $|\til \psi_{\til{\b p}}|^2 \propto \theta(v_0/v - |\til{\b p}|)$, may also be used. However, the formation of dark matter halos and solitons is largely insensitive to the initial distribution \cite{Levkov:2018kau}. To study isolated halos, we run simulations in a box with size $\til L > 2\pi/\til k_J$, where $\til k_J = (4\til{\bar\rho})^{1/4}$ is the Jeans scale, with a bar over variables indicating their spatial mean value. The box size is set to $\til{L} = 18$ and the total number of particles to $\til{N} = 1005.3$.

\section{Formation and collapse of solitons in dark matter halos}
\label{sec:formation}
In this section, we examine how solitons form, grow and collapse in dark matter halos influenced by NGIs. We begin by reviewing the soliton formation process in the absence of NGIs (i.e., $\til{\xi} = 0$) in section \ref{sec:no_ngi}. Next, we outline some key properties of solitons in section \ref{sec:soliton_ngi}. We then analyze the impact of NGIs with positive and negative coupling constants $\til\xi$ in sections \ref{sec:pos_ngi} and \ref{sec:neg_ngi}, respectively.

\subsection{Condensation of self-gravitating solitons}
\label{sec:no_ngi}
For self-gravitating dark matter without NGIs, studies have demonstrated that gravity induces the formation of solitons at the centers of dark matter halos. The condensation timescale for dark matter particles within a halo of size $R_h$ is given by \cite{Levkov:2018kau}
\begin{align}
\label{soliton_condensation_time}
t_\rm{gr} = \frac{16 \sqrt{2} b}{3\pi} \frac{\MP^4mv_0^6}{n^2 \Lambda} ~,
\end{align}
where $b$ is an $\cal O(1)$ coefficient determined by simulations, $v_0$ is the characteristic velocity, $n$ is the typical number density, and $\Lambda=\log(mv_0 R_h)$ is the Coulomb logarithm. Once formed, solitons continue to grow by accreting mass from the surrounding gas of particles at a rate $M_s \propto t^{1/2}$ \cite{Levkov:2018kau}. As the soliton grows sufficiently large, it begins to heat the surrounding halo and makes the whole system reach virial equilibrium after a few condensation times $t_\rm{gr}$ \cite{Eggemeier:2019jsu}. The mass growth of solitons is then slowed down to a rate $M_s\propto t^{1/8}$ \cite{Eggemeier:2019jsu, PhysRevD.104.083022} or $t^{1/4}$ \cite{Chan:2022bkz}. These findings present conflicting results; however, recent work suggests that the soliton mass evolves not as a simple power law over time, potentially encompassing both of these behaviors \cite{Dmitriev:2023ipv}. This paper does not delve into resolving this debate.

\begin{figure}
\centering
\includegraphics[width=\linewidth]{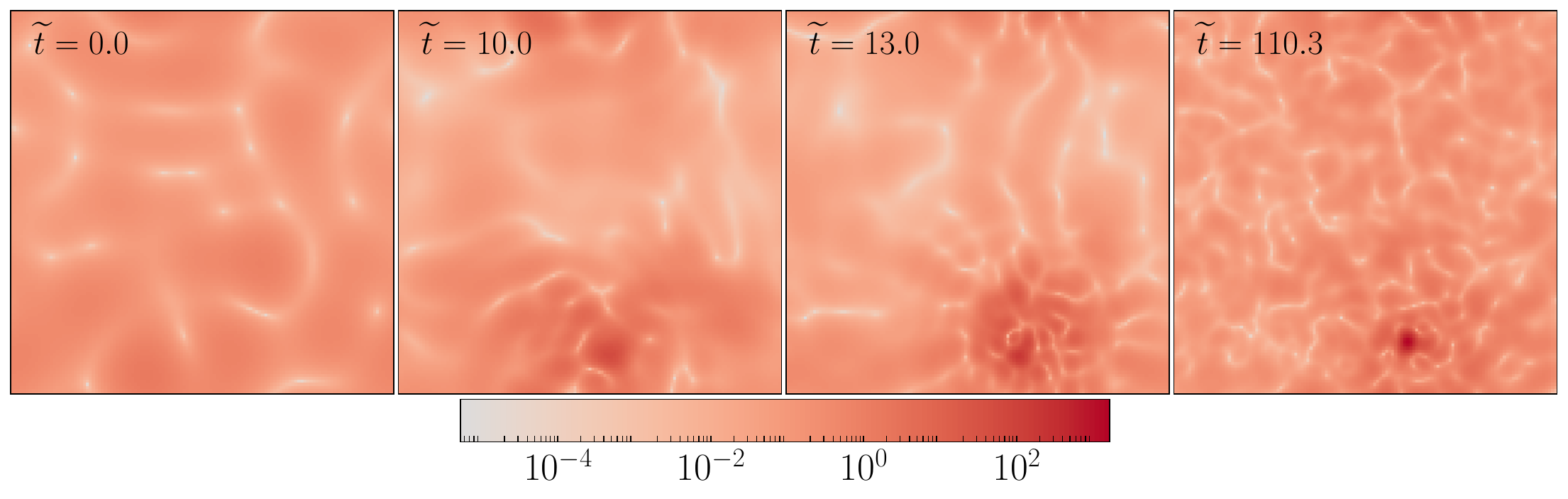}
\caption{Evolution of dark matter without nonminimal gravitational effects, illustrated by slices of the particle number density $\til\rho$ at different times. These slices show the density on a plane passing through the peak density location. We see that a dense dark matter halo forms gradually over time, with a soliton clearly emerging at the halo's center by $\til t\approx110$.}
\label{fig:rhoxi0}
\end{figure}
For completeness, we simulate dark matter without NGIs and present the structure formation in figure \ref{fig:rhoxi0}. We observe the gradual formation of a dense dark matter halo from $\til{t} = 0$ to $\til{t} = 10$. By $\til{t} \approx 110$, a distinct dense object, with a density profile closely resembling a soliton solution, becomes evident at the halo's center, in agreement with the soliton condensation timescale given by \eqref{soliton_condensation_time}.

\subsection{Nonminimally gravitating solitons}
\label{sec:soliton_ngi}
The natural emergence of solitons in dark matter halos can be attributed to their status as ground states of the Schroedinger-Poisson equations at fixed particle number \cite{Lee:1991ax, Hui:2016ltb, Zhang:2023ktk}, making them energetically favorable in a thermal bath of nonrelativistic particles. In this subsection, we provide a concise overview of key properties of solitons in the presence of NGIs, following \cite{Zhang:2024bjo}.

Generally speaking, scalar solitons take the form $\til \psi(\til t,\til{\b x}) = \til f(\til r) e^{i \til\mu \til t}$, where $\til \mu$ is the chemical potential and $\til f$ represents a spherically symmetric profile. The field equations \eqref{eom3} and \eqref{eom4} then translate into radial profile equations for $\til f$ and $\til\Phi$:
\begin{align}
\label{soliton_profile}
-\frac{1}{2}\til\nabla^2\til f + (\til\Phi+\til\mu) \til f + \til\xi \til f^3 + \til\xi^2 \til f \til\nabla^2\til f^2 = 0 ~,
\end{align}
and $\til\nabla^2\til\Phi = \til f^2 + \til\xi \til\nabla^2 \til f^2$. Soliton solutions can then be found by taking the boundary conditions $\pd_{\til r}\til f|_{\til r=0} = \pd_{\til r}\til \Phi|_{\til r=0}=0$ and $\pd_{\til r}\til f|_{\til r\rightarrow \infty} = 0$. Some examples are shown as solid curves in figures \ref{fig:profilexi0.02} and \ref{fig:profilexi-0.008}, obtained using the Mathematica package DMSolitonFinder developed by one of us \cite{Zhang:2024bjo}.

The term $\til\xi\til f^3$ in \eqref{soliton_profile} plays a similar role to a quartic self-coupling of the progenitor Lorentz scalar field. A positive $\til\xi$ tends to compactify solitons, requiring more matter to balance the induced repulsive force. A negative $\til\xi$ destabilizes solitons as their mass increases, reaching a critical density where the induced attractive force competes with gravity. At the critical densities,
\begin{align}
\label{max_rhoxi_negative}
\til\rho^\rm{max} = 0.0458 |\til\xi|^{-2}
\quad\text{and}\quad
\til\rho_\xi^\rm{max} = 0.0884 |\til\xi|^{-2} \quad(\text{for } \til\xi<0) ~,
\end{align}
solitons collapse until the relativistic theory is warranted to depict the full story.

The impact of the gradient-dependent part of NGIs in equation \eqref{soliton_profile}, $\til\xi^2 \til f\til\nabla^2 \til f^2$, is more complex. A positive $\til\xi$ makes the mass density $\til\rho_\xi$ less than the particle number density $\til\rho$ and thus induces a repulsive force within soliton cores, with causing opposite effects in the outer regions. It also imposes an upper limit on central soliton amplitudes, $\til f_0^\rm{max} = |2\til\xi|^{-1}$. This corresponds to the maximum densities
\begin{align}
\label{max_rhoxi_positive}
\til\rho^\rm{max} = 0.250 \til\xi^{-2}
\quad\text{and}\quad
\til \rho_\xi^\rm{max} = 0.0746 \til\xi^{-2} \quad(\text{for } \til\xi>0) ~.
\end{align}
In such critical solitons, the repulsive force is sufficiently strong that the core may exhibit negative density $\rho_\xi$, as we will see in next subsection. Similar amplitude and density bounds exist for negative $\tilde{\xi}$, which nevertheless remain on the unstable branch of solitons and thus do not have significant impact.

\subsection{Dark matter with positive nonminimal couplings}
\label{sec:pos_ngi}
To explore the formation of dark matter halos and solitons under the influence of NGIs in an efficient way, we need to determine an appropriate magnitude for $\til\xi$. A useful quantity for assessing the importance of NGIs is the Jeans scale, which govern the scales of linearized perturbations that experience rapid growth. For $\til\xi \sqrt{\til\rho}<1$, the Jeans scale is approximately given by $\til k_J \approx (4\til{\rho})^{1/4} [ 1+\til\xi \sqrt{ \til{\rho} } ]$ \cite{Zhang:2023fhs}, indicating that NGIs may only be relevant if $\til\xi\sqrt{\til\rho} \gtrsim 1$. However, nonlinear analysis of solitons suggests that NGIs start to significantly affect dynamics around $\til\xi\sqrt{\til\rho} \sim 0.3$, as indicated by \eqref{max_rhoxi_negative} and \eqref{max_rhoxi_positive}. Therefore, we adopt $|\til\xi| \sim \cal O(0.01)$ for our simulations. At these values of $|\til\xi|$, the initial evolution of dark matter resembles that without NGIs and the condensation time scale \eqref{soliton_condensation_time} will not be significantly affected, but the impact of NGIs becomes pronounced at later times $\til t\sim \cal O(10)$ as inhomogeneities develop.

\begin{figure}
\centering
\includegraphics[width=\linewidth]{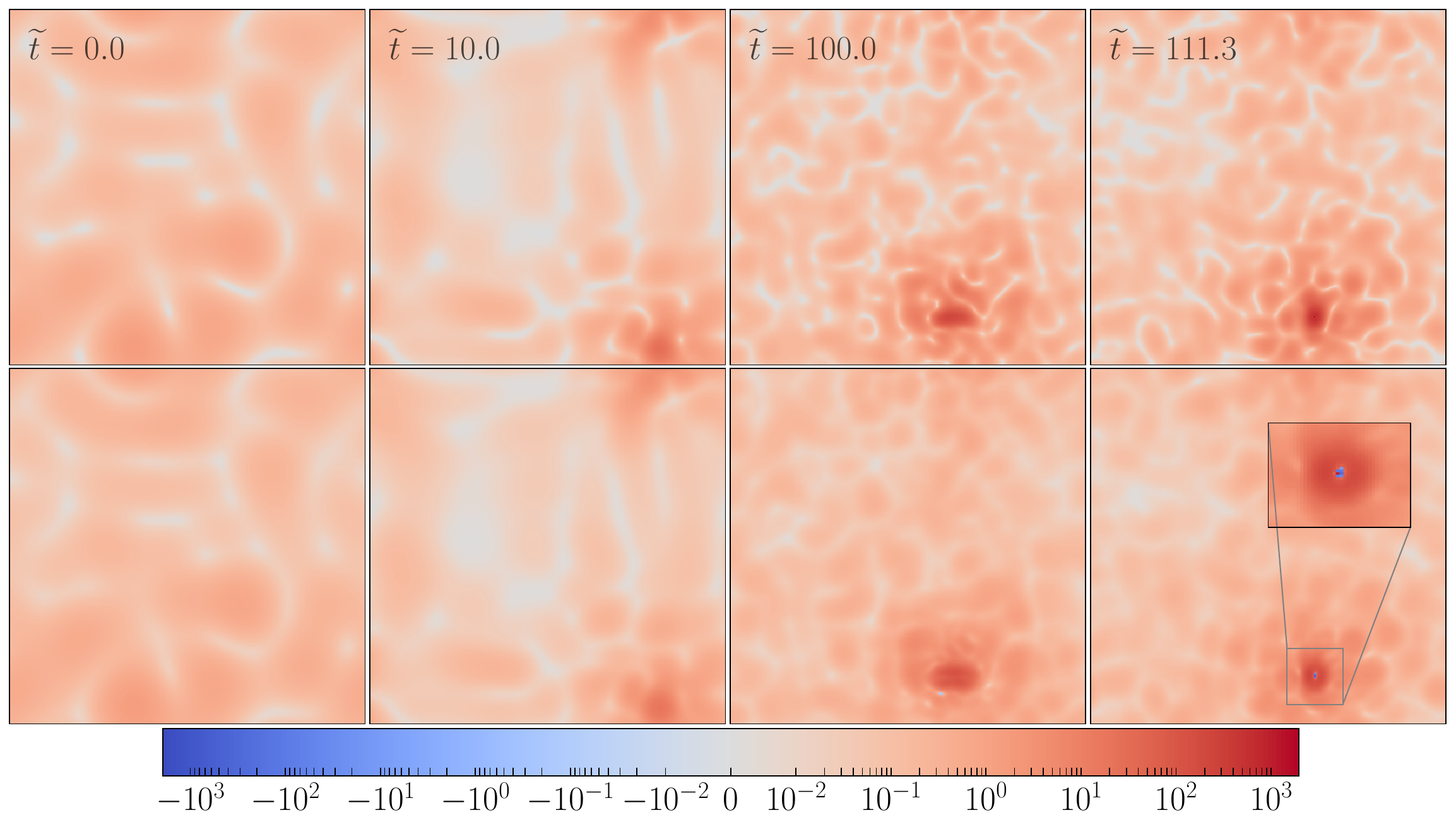}
\caption{Evolution of dark matter with $\til\xi=0.02$, illustrated by slices of the number density $\til\rho$ (in upper panels) and the mass density $\til\rho_\xi$ (in lower panels) at different times. By $\til t=100$, a soliton has formed at the halo's center. As the soliton grows, the mass density $\til\rho_\xi$ in its inner core becomes negative when the soliton collapses at $\til t\approx 111.3$.}
\label{fig:rhoxi0.02}
\end{figure}

In figure \ref{fig:rhoxi0.02}, we show the evolution of dark matter densities $\til\rho$ (in upper panels) and $\til\rho_\xi$ (in lower panels) with $\til\xi=0.02$. At early times $\til t\lesssim 10$, a dark matter halo gradually forms, during which the distributions of $\til\rho$ and $\til\rho_\xi$ look quite similar. Their difference, $\til\rho_\xi - \til\rho = \til\xi\til\nabla^2 \til\rho$, remains suppressed since inhomogeneities are not sufficiently developed. 

\begin{figure}
\centering
\begin{minipage}{0.49\textwidth}
\includegraphics[width=\linewidth]{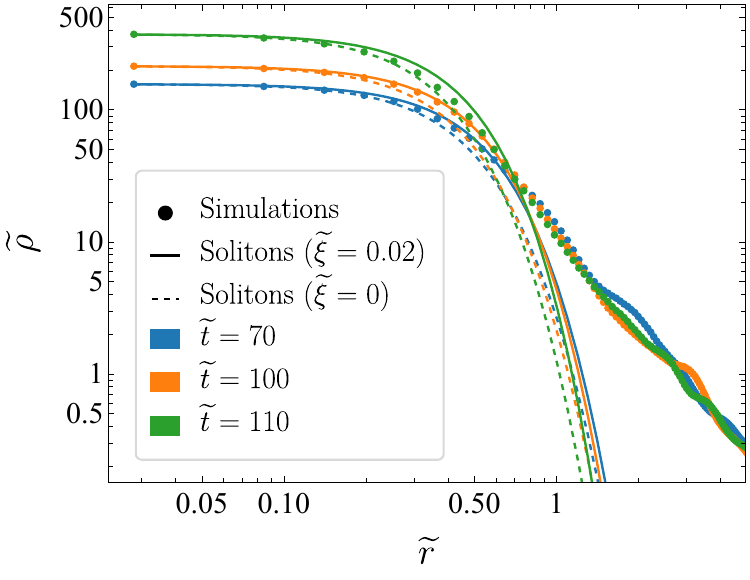}
\end{minipage}\hfill
\begin{minipage}{0.49\textwidth}
\includegraphics[width=\linewidth]{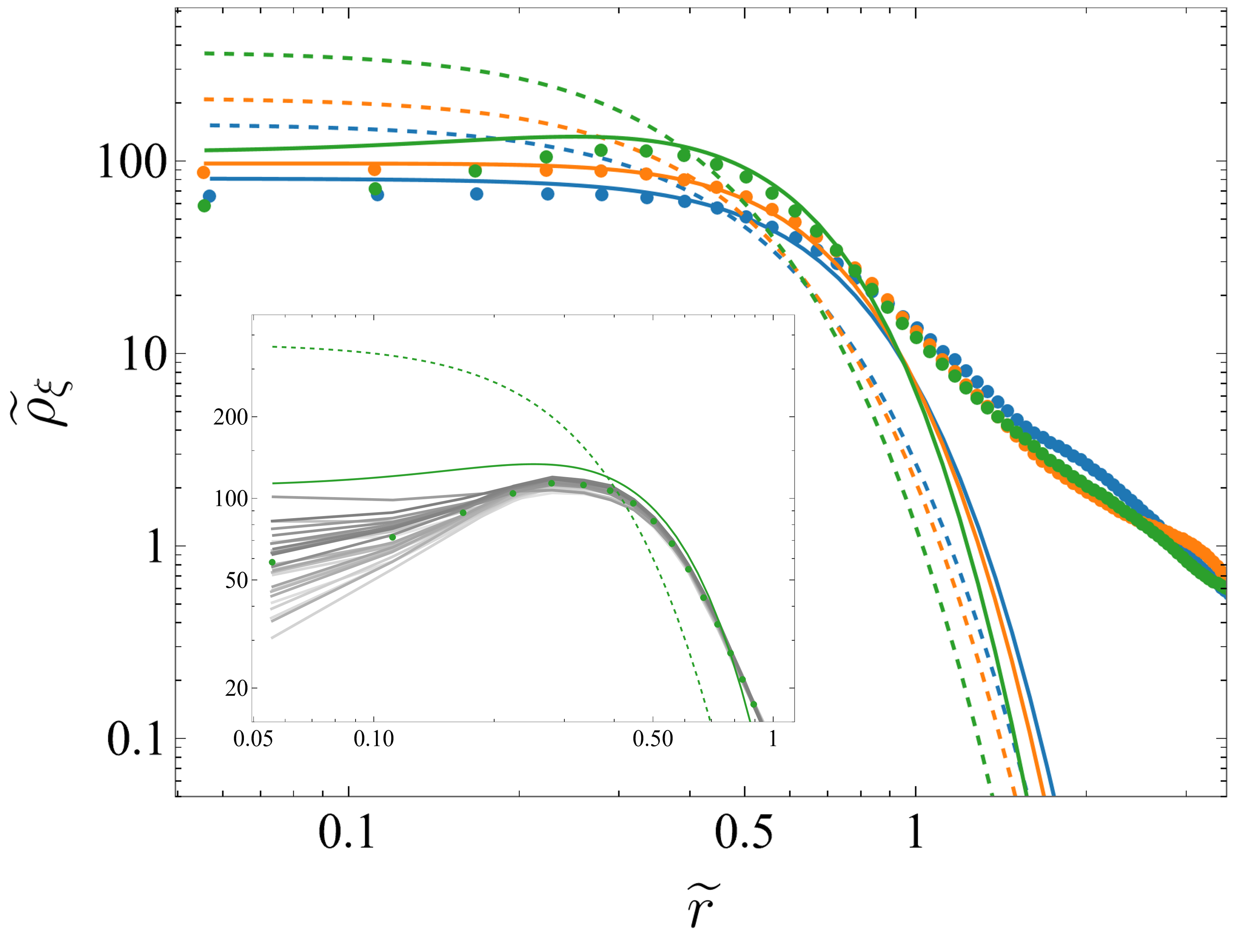}
\end{minipage}
\caption{Density profiles of the halo at different times. The data points are taken from simulations with $\til\xi=0.02$, and are well fitted by the soliton solution with an identical number density $\til\rho$ at the center (solid lines). Soliton profiles without NGIs are also shown for comparison (dashed lines). For the numerical data points of $\til\rho_\xi$, we average 27 profiles (gray lines in the inset, obtained at $\til t=110$ for illustration) extracted by considering slightly different locations as the halo's center to mitigate numerical errors.}
\label{fig:profilexi0.02}
\end{figure}

By $\til t=100$, the center of the halo becomes very dense, indicating the presence of a soliton. This is confirmed by comparing density profiles of the halo and the soliton solution with the same particle number density $\til\rho$ at the center. Figure \ref{fig:profilexi0.02} shows the density profiles of the halo and the corresponding soliton solution at various times (dots and solid lines, respectively). As matter accretes, the $\til\rho_\xi$ profile at small radii becomes increasingly sensitive to the selection of the halo's center. The inset of the right plot displays 27 profiles extracted at $\til t=110$ by taking slightly different locations around the maximum $\til\rho$ as the halo's center (gray lines). They differ at $\til r\lesssim 2|\til\xi|^{1/2}$ but converge at larger radii. To mitigate this uncertainty, we average these 27 profiles to obtain numerical data points solely for $\til\rho_\xi$ with $\til\xi=0.02$. Both $\til\rho$ and $\til\rho_\xi$ profiles show that the halo and soliton solution coincide well at $\til r\lesssim 5 |\til\xi|^{1/2}$. In contrast, solitons with $\til\xi=0$ (dashed lines) poorly fit the numerical data.

\begin{figure}
\centering
\begin{minipage}{0.49\textwidth}
\includegraphics[width=\linewidth]{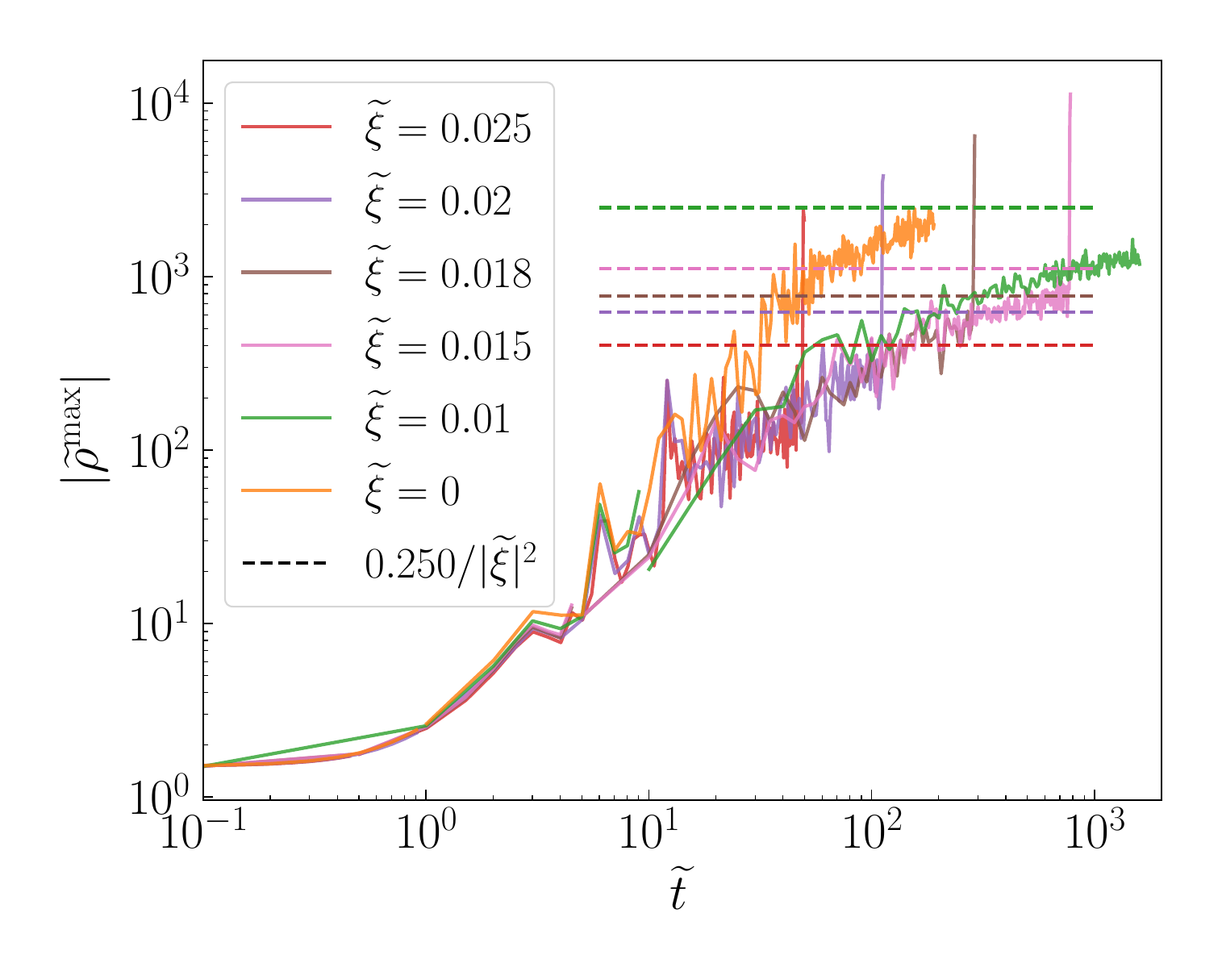}
\end{minipage}\hfill
\begin{minipage}{0.49\textwidth}
\includegraphics[width=\linewidth]{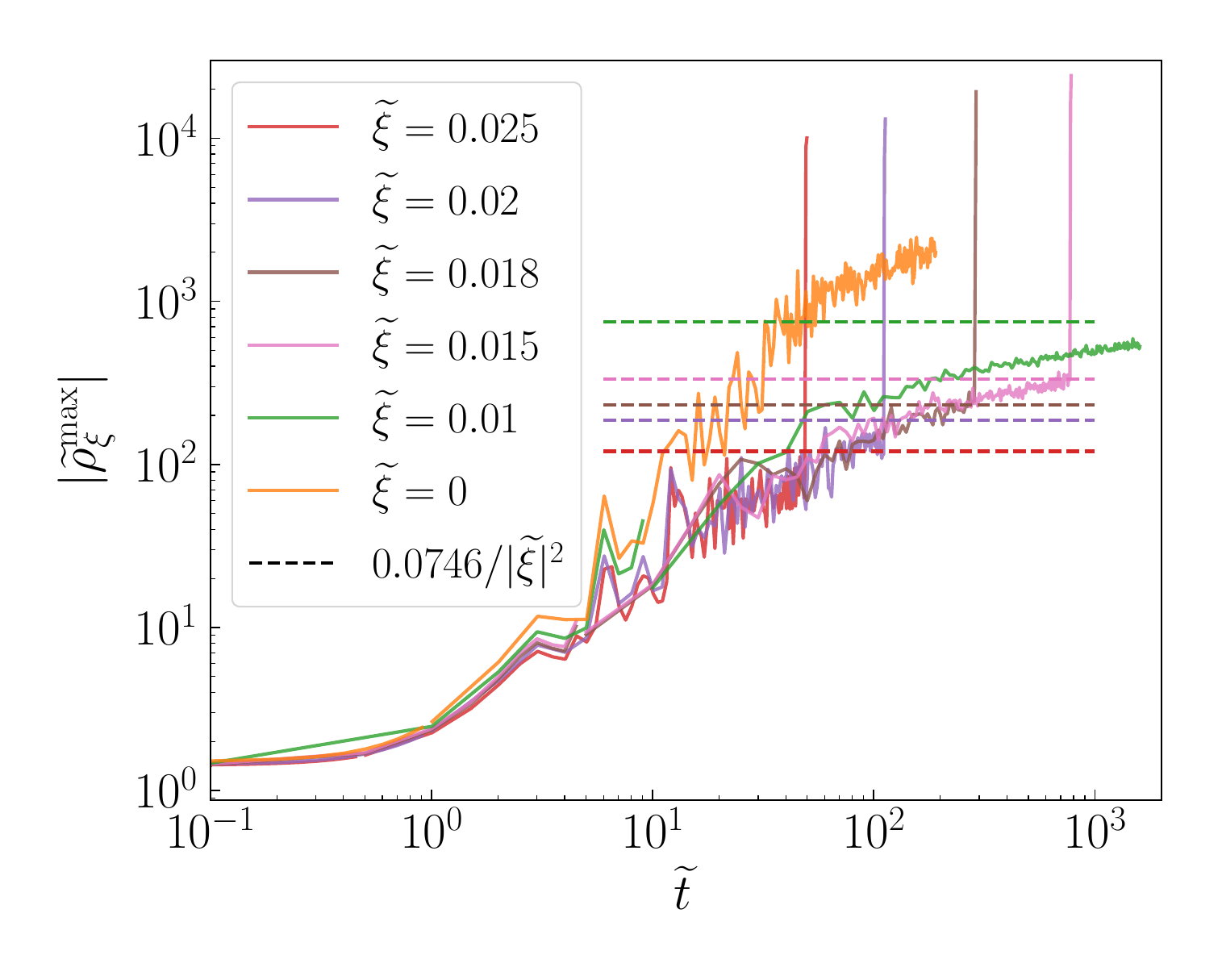}
\end{minipage}
\caption{Evolution of the maximum number density $\til\rho^\rm{max}$ (left) and the maximum mass density $\til\rho_\xi^\rm{max}$ (right) for various positive NGI couplings. Dashed lines indicate the maximum densities predicted by \eqref{max_rhoxi_positive}, beyond which solitons are expected to collapse.}
\label{fig:maxrhoxipositive}
\end{figure}

The soliton keeps growing after formation and is expected to collapse when its densities reach the critical values \eqref{max_rhoxi_positive}. This prediction is verified through multiple simulations with various values of $\til\xi$. In figure \ref{fig:maxrhoxipositive}, the evolution of the maximum density $\til\rho_\xi^\rm{max}$ is represented by solid lines, while dashed lines indicate the corresponding critical densities. At collapse, the soliton radius, defined as the radius where the particle number density $\til\rho$ drops to half of its central value, is given by
\begin{align}
R_s = 1.36 \times 10^{8} \rm{km} \( \frac{\xi}{10} \) \( \frac{10^{-18}\rm{eV}}{m} \) ~.
\end{align}

Due to the gradient-independent part of NGIs, one may expect a regime where gravity is counterbalanced by repulsive NGIs, with negligible quantum pressure, allowing solitons to be well described by the Thomas-Fermi approximation \cite{PhysRevD.104.083022}. However, this regime lies beyond the critical densities \eqref{max_rhoxi_positive} and cannot be achieved dynamically via gravitational accretion.

\subsection{Dark matter with negative nonminimal couplings}
\label{sec:neg_ngi}

\begin{figure}
\centering
\includegraphics[width=\linewidth]{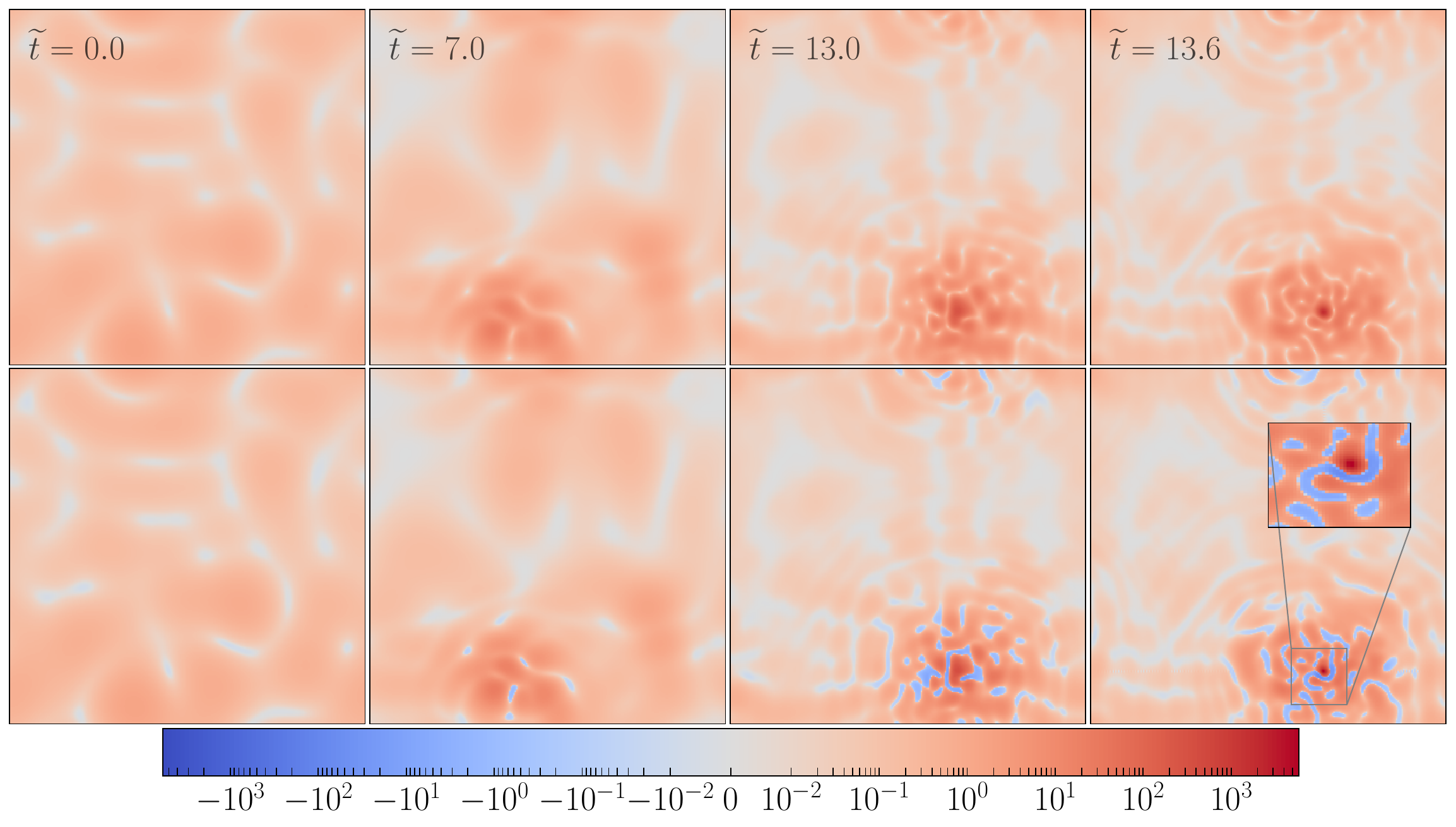}
\caption{Evolution of dark matter with $\til\xi=-0.008$, illustrated by slices of the number density $\til\rho$ (in upper panels) and the mass density $\til\rho_\xi$ (in lower panels) at different times. By $\til t=13$, a soliton has formed at the center of the halo, surrounded by multiple subhalos. These subhalos could contain more mass than particle numbers, with regions showing negative mass density $\til\rho_\xi$ in the voids between them.}
\label{fig:rhoxi-0.008}
\end{figure}

\begin{figure}
\centering
\begin{minipage}{0.49\textwidth}
\includegraphics[width=\linewidth]{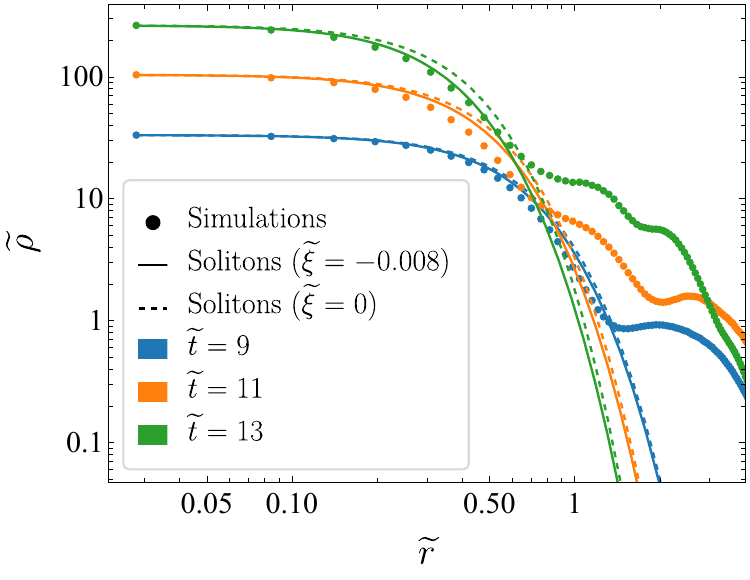}
\end{minipage}\hfill
\begin{minipage}{0.49\textwidth}
\includegraphics[width=\linewidth]{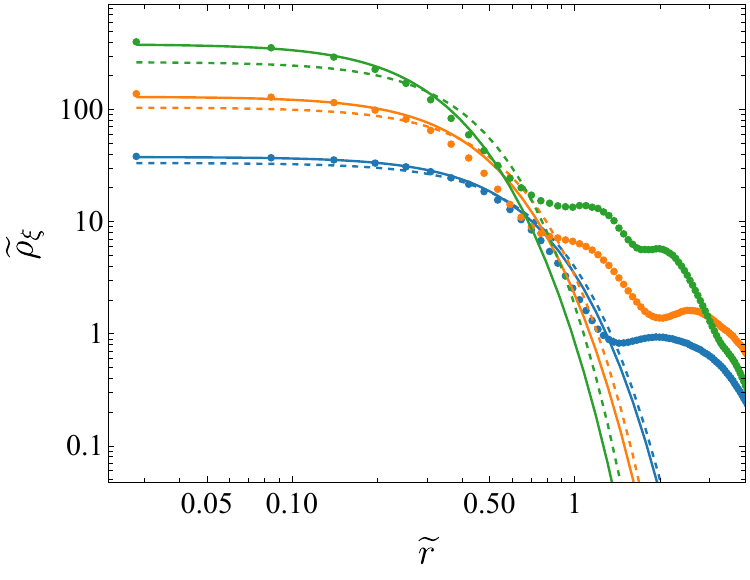}
\end{minipage}
\caption{Density profiles of the halo at different times. The data points are taken from simulations with $\til\xi=-0.008$, and are well fitted by the soliton solution with an identical number density $\til\rho$ at the center (solid lines). Soliton profiles without NGIs are also shown for comparison (dashed lines).}
\label{fig:profilexi-0.008}
\end{figure}

In this subsection, we examine the evolution of dark matter with negative $\til\xi$. Figure \ref{fig:rhoxi-0.008} illustrates the evolution of dark matter densities $\til\rho$ (upper panels) and $\til\rho_\xi$ (lower panels) with $\til\xi=-0.008$. Similar to the case with positive couplings, we observe that the difference between $\til\rho$ and $\til\rho_\xi$ is small during the initial stages of evolution. By $\til t=13$, a cluster of dark matter halos has formed, containing a central soliton surrounded by many subhalos. The profiles of the soliton are depicted in figure \ref{fig:profilexi-0.008}. The subhalos typically contain higher mass than particle numbers, with regions exhibiting negative mass density $\til\rho_\xi$ in the voids between them.

The evolution of maximum densities are shown in figure \ref{fig:rhoxi-0.008}. The soliton continues to grow after formation and eventually collapses when its maximum densities predicted by \eqref{max_rhoxi_negative} (indicated by the dashed lines), as expected.  Above the critical values, solitons are no longer energetically favored states and become unstable to perturbations. At collapse, the radius of the soliton, where the particle number density drops to half of its central value, is given by
\begin{align}
R_s = 4.32 \times 10^8 \rm{km} \( \frac{|\xi|}{10} \) \( \frac{10^{-18}\rm{eV}}{m} \) ~.
\end{align}
As mentioned in section \ref{sec:soliton_ngi}, there exist additional upper limits on soliton densities due to the gradient-dependent part of NGIs. However, these densities correspond to the unstable branch and are not accessible in our simulations.

\begin{figure}
\centering
\begin{minipage}{0.49\textwidth}
\includegraphics[width=\linewidth]{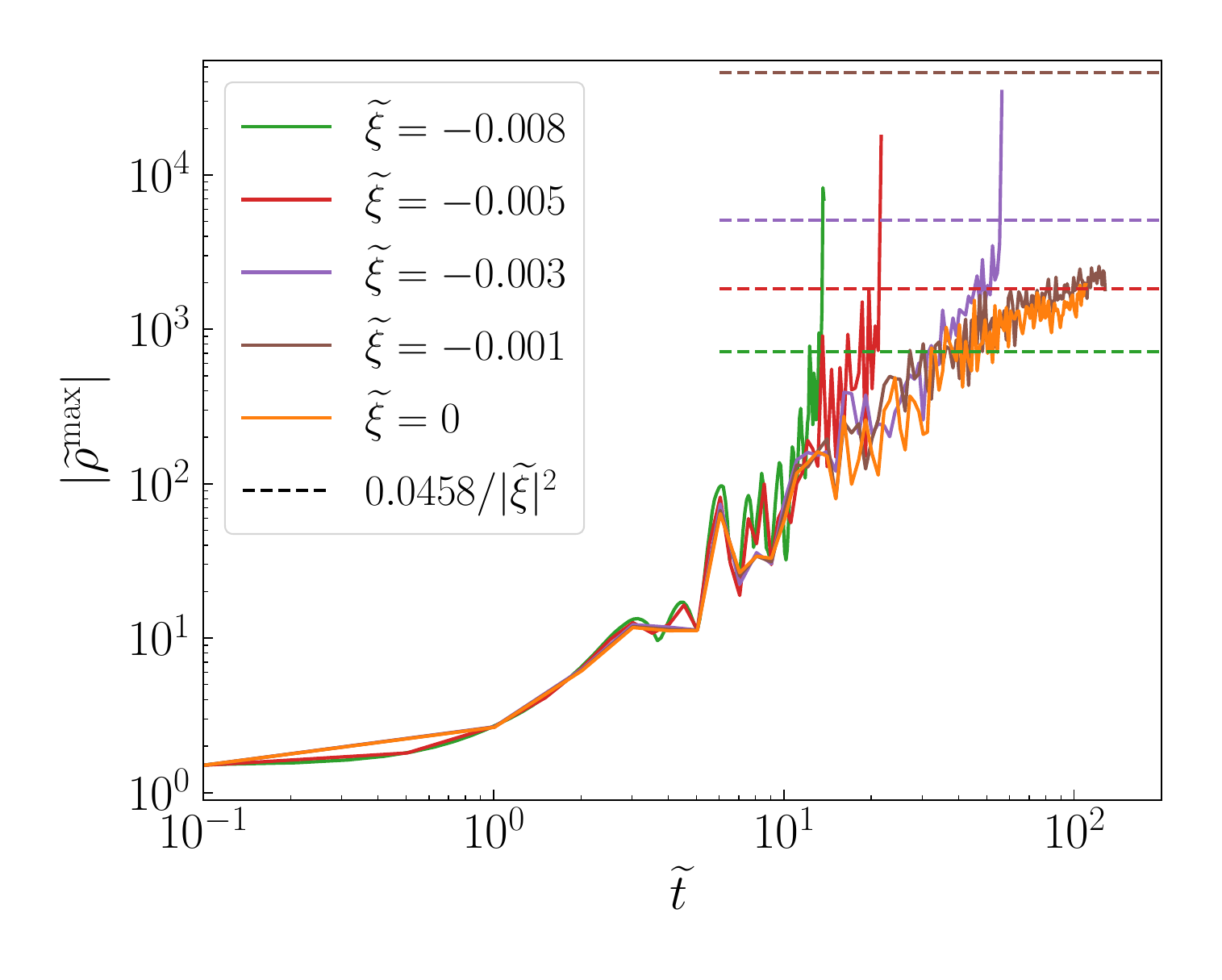}
\end{minipage}\hfill
\begin{minipage}{0.49\textwidth}
\includegraphics[width=\linewidth]{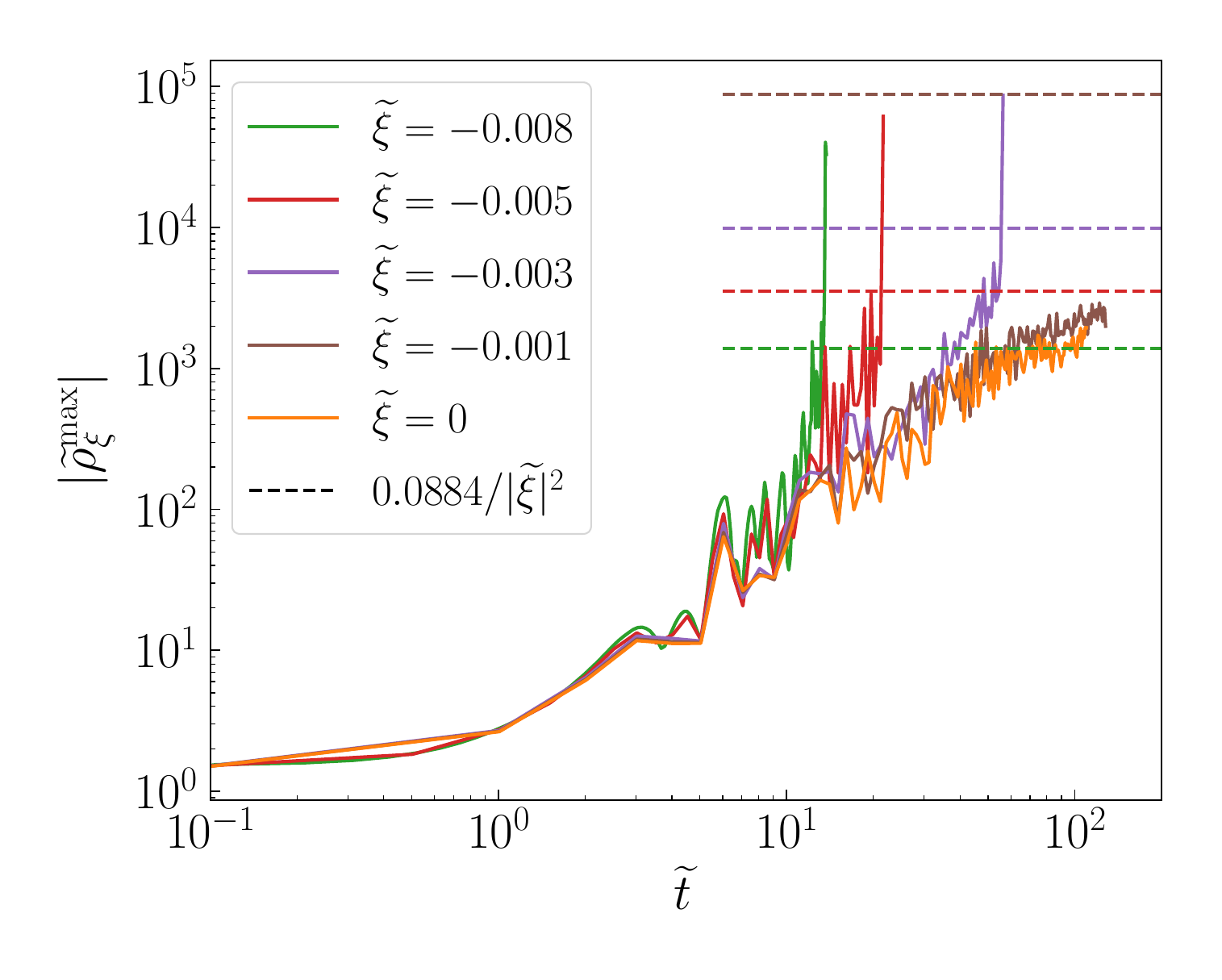}
\end{minipage}
\caption{Evolution of the maximum number density $\til\rho^\rm{max}$ (left) and the maximum mass density $\til\rho_\xi^\rm{max}$ (right) for various negative NGI couplings. Dashed lines indicate the maximum densities predicted by \eqref{max_rhoxi_negative}, beyond which solitons are expected to collapse.}
\label{fig:maxrhoxinegative}
\end{figure}

\section{Conclusions}
\label{sec:conclusion}
In this work, we study the formation, growth, and collapse of solitons within halos of nonminimally gravitating dark matter, characterized by both gradient-independent and gradient-dependent self-interactions in the nonrelativistic regime. This latter component, despite being overlooked in parameterizations for self-interacting ultralight dark matter, could have important implications for the phenomenology of dark matter and its solitons. We investigate the effects by numerically evolving the nonlinear Schroedinger-Poisson equations \eqref{eom3} and \eqref{eom4}. Although our simulations are not cosmological, our findings are relevant in cosmological models possessing appropriate environments.

In figures \ref{fig:maxrhoxipositive} and \ref{fig:maxrhoxinegative}, we demonstrate that for both positive and negative coupling constants $\til\xi$, solitons collapse when their densities exceed certain critical values given by equations \eqref{max_rhoxi_negative} and \eqref{max_rhoxi_positive}. The mass of collapsing solitons is at the level of $\sim 40\MP^2/(m|\xi|^{1/2})$ \cite{Zhang:2024bjo} . Once the critical densities are reached, the dark matter clump rapidly grows, necessitating relativistic simulations for further evolution. In contrast, solitons with solely gradient-independent self-interactions will collapse only for negative couplings, i.e., attractive self-interactions \cite{PhysRevD.104.083022, PhysRevD.106.023009, Mocz:2023adf}. Collapsing solitons undergo a series of bursts and could be potentially detected if dark matter is coupled to regular matter \cite{Levkov:2016rkk}.

Another novel aspect of nonminimally gravitating dark matter is the inclusion of a gradient-dependent source in the Poisson's equation \eqref{eom2}, alongside the rest mass density. This leads us to consider a modified mass density $\rho_\xi$ that maintains the familiar Newton's law. Our numerical results suggest that solitons with positive couplings could feature an inner core with negative mass density via dynamical matter accretion, as shown in figure \ref{fig:rhoxi0.02}. For negative couplings, subhalos surrounding the central soliton could have more mass than particle numbers with voids of negative mass density in between, see figure \ref{fig:rhoxi-0.008}. These characteristics could have novel implications for gravitational lensing \cite{Fujikura:2021omw}, dynamic heating of stars \cite{Dalal:2022rmp}, and axion minivoids \cite{Eggemeier:2022hqa}, warranting further investigation.

We also simulate systems with stronger nonminimal couplings of the order $\til\xi\sqrt{\til\rho} \sim \cal O(0.1)$. In these cases, structure formation is significantly enhanced and density spikes rapidly develop beyond our numerical resolutions. The fast formation can be understood by noting that linearized density perturbations could grow on both large and small scales if the nonminimal coupling is strong enough that $\til\xi\sqrt{\til\rho} \sim 1$ \cite{Zhang:2023fhs}. We verify this novel pattern by limiting the simulation box to below the usual Jeans scale, i.e., $\til L < 2\pi/\til k_J$. We observe the development of mini structures, though our simulations quickly break down due to amplified noise.

In conclusion, our study reveals novel insights into the formation and growth of dark matter halos and solitons influenced by nonminimal gravitational effects. This research opens up new avenues for probing ultralight dark matter and modified gravity through cosmological and astrophysical observations.

\acknowledgments
We would like to thank Xiaolong Du and Luca Visinelli for helpful comments. We would especially like to thank Zhipan Li for his assistance in setting up the clusters. JC acknowledges the support from the Fundamental Research Funds for the Central Universities (SWU-KR22012) and the Chongqing Natural Science Foundation General Project (2023NSCQ-MSX1929). HYZ is supported in part by the National Natural Science Foundation of China (NSFC) through grant No. 12350610240. This research was also facilitated by the computational resources in the School of Physical Science and Technology at Southwest University.

\bibliographystyle{jhep}
\bibliography{ref}
\end{document}